# Observational Constraints on the Great Filter

Jacob Haqq-Misra,[1,*] Ravi Kumar Kopparapu,[2] and Edward Schwieterman[1,3]

[1]Blue Marble Space Institute of Science, 1001 4th Ave, Suite 3201, Seattle WA 98154, USA.
[2]NASA Goddard Space Flight Center, Greenbelt MD, USA
[3]University of California at Riverside, Riverside CA, USA
[*]Email: jacob@bmsis.org

**Abstract**
The search for spectroscopic biosignatures with the next-generation of space telescopes could provide observational constraints on the abundance of exoplanets with signs of life. An extension of this spectroscopic characterization of exoplanets is the search for observational evidence of technology, known as technosignatures. Current mission concepts that would observe biosignatures from ultraviolet to near-infrared wavelengths could place upper limits on the fraction of planets in the galaxy that host life, although such missions tend to have relatively limited capabilities of constraining the prevalence of technosignatures at mid-infrared wavelengths. Yet searching for technosignatures alongside biosignatures would provide important knowledge about the future of our civilization. If planets with technosignatures are abundant, then we can increase our confidence that the hardest step in planetary evolution—the Great Filter—is probably in our past. But if we find that life is commonplace while technosignatures are absent, then this would increase the likelihood that the Great Filter awaits to challenge us in the future.

## 1  Introduction

NASA is engaged in a systematic effort at characterizing the spectral signatures of exoplanets through ground-based and space-based observations. One compelling long-term objective is to search for spectroscopic biosignatures on exoplanets that would indicate the presence of extraterrestrial biospheres. The simultaneous detection of water vapor, oxygen, ozone, carbon dioxide, and methane is an example of a spectroscopic biosignature based on Earth life over time, with many significant spectral features from ultraviolet to near-infrared wavelengths (Des Marais et al. 2002, Arney et al. 2016, 2018; Schwieterman et al. 2018; Krissansen-Totton et al. 2018). Another example of a remotely detectable biosignature is the "red edge" near-infrared reflectance spectrum from large-scale photosynthesis (e.g., Seager et al. 2005; Turnbull et al. 2006; Kiang et al. 2007). A substantial and ongoing investigation is focused on identifying plausible biosignatures and anti-biosignatures to prepare for such observations (e.g., Seager et al. 2012; Grenfell 2017; Kaltenegger 2017; Catling et al. 2018; Fujii et al. 2018; Kiang et al. 2018; Meadows et al. 2018; Schwieterman et al. 2018; Walker et al. 2018; Lammer et al. 2019; O'Malley & Kaltenegger 2019).

Similar to biosignatures, spectroscopic technosignatures would indicate the presence of extraterrestrial technology on exoplanets. The search for technosignatures, by analogy with biosignatures, refers to any detectable signs of a planetary system that has been modified by technology (Wright & Gelino 2018). Technosignatures logically share a continuum with the search for biosignatures, both of which span a broad range of possibilities. Chlorofluorocarbons (CFCs), perfluorocarbons (PFCs), and other artificial greenhouse gases are atmospheric constituents on Earth due to technology alone, with prominent mid-infrared absorption features that provide examples of spectroscopic technosignatures (Marinova et al. 2005; Schneider et al. 2010; Lin et al. 2014). Large, complex molecules that are technologically manufactured generally show a larger number of rotovibration transitions at mid-infrared wavelengths, particularly in the relatively low-temperature environment of a habitable planet (able to maintain surface liquid water). Another possible technosignature is excess mid-infrared emission from a planetary system that would indicate "waste heat" from large-scale technological structures (e.g., Wright et al. 2014ab, 2015; Griffith 2015). Surface properties could also serve as technosignatures, such as artificial spectral edges analogous to the vegetation red-edge but at wavelengths that would be expected from a civilization with planetary-scale photovoltaics (Lingam & Loeb 2017). A handful of studies have begun to explore the possibility for observing technosignatures and anti-technosignatures (e.g., Whitmire & Wright 1980; Loeb & Turner 2012; Lin et al. 2014; Kuhn & Berdyugina 2015; Kipping & Teachey 2016; Stevens et al. 2016; Lingam & Loeb 2017; Wright & Gelino 2018; Pinchuck et al. 2019), but the field of planetary spectroscopic technosignatures is in its infancy compared to the science of biosignatures.

The next generation of space telescopes currently being studied by NASA are capable of observing biosignatures on terrestrial planets orbiting within the habitable zone of their host star. Current candidates for this mission include the Large Ultraviolet Optical Infrared Surveyor (LUVOIR), which would search for spectral biosignatures in the wavelength range of 200 to 2000 nm using a mirror 8 to 15 meters in diameter (Fischer et al. 2019). The 15-meter LUVOIR architecture would be capable of characterizing ~60 rocky habitable zone planets during its mission duration, while the 8-meter version could characterize ~30-40 rocky habitable zone planets (Fischer et al. 2019). The Habitable Exoplanet (HabEx) imaging mission is also optimized for identifying spectral biosignatures from the ultraviolet to near-infrared using a 4-meter telescope (Gaudi et al. 2019) and could characterize ~10 rocky habitable zone planets. (It is important to note that rocky planets in the habitable zone are not necessarily habitable). The Origins Space Telescope (OST) mission is intended for mid-to-far-infrared astronomy and could detect spectral features of ozone as well as possible infrared technosignatures (Cooray et al. 2019). The advantage of OST is that it can detect transit spectra of planetary systems at a greater maximum range than LUVOIR, but only for late-type (M dwarf) hosts with transiting planets. The final mission selection and specifications remain to be decided, but it is within technical capabilities for such a mission to characterize a large number of terrestrial planets—although the overall yields are higher for LUVOIR than HabEx or OST. If biospheres like Earth are commonplace, then such missions could reveal a statistically significant sample of biosignatures that observationally constrain the frequency of inhabited planets in the galaxy.

## 2 The Great Filter

A prominent assumption among scientists and non-scientists alike is that any extraterrestrial technological civilizations would have attempted to expand across the galaxy. The peculiar absence of evidence for settlement of the galaxy or visitation of the Solar System suggests that such expansionistic extraterrestrial civilizations may not exist. This argument has taken various forms (e.g., Hart 1975; Tipler 1980) and is often described as the Fermi paradox (Gray 2015) or the Great Silence (Brin 1983). One resolution to the Great Silence is that life or civilization on Earth is extremely rare in the universe, which suggests that humans are unlikely to ever discover signs of another technological civilization (e.g., Ward & Brownlee 2000). However, the Great Silence argument has instigated a broad range of resolutions, many of which are compatible with a universe teeming with intelligent life that still appears empty to humans today (Webb 2015; Cirkovic 2018; Forgan 2019).

Resolving the Great Silence also carries implications for the future of technological civilization on Earth. The apparent deadness of nearby space and absence of evidence for extraterrestrial technology suggest that there is at least one extremely improbable evolutionary step somewhere from the origin of life to galactic-scale settlement. Hanson (1998) called this argument the "Great Filter" as a descriptor of the inhibitory step in evolution, which then raises "the ominous question: how far along this filter are we?" If the origin of life is an improbable event, then this would explain the lack of abundant life on nearby planets, with the Great Filter in our past. But Hanson (1998) also pointed out that "evidence of extraterrestrials is likely bad (though valuable) news" because this would mean the Great Filter awaits in our future. If observations show the galaxy to be teeming with complex life, and even Earth-like civilizations, then this would suggest that the evolutionary steps from the origin of life up until today have been relatively benign. This means that the Great Filter is in our future, somewhere after our current technological state but before galactic-scale settlement. The irony of this reasoning is that the discovery of extraterrestrial life, a seemingly monumental discovery, would also imply a bleaker future for humanity.

The Great Filter even carries implications for the search for life in the Solar System. The discovery of extinct or extant life on any other planet would indicate that at least one of the difficult steps in evolution are beyond the origin of life and closer to today. This reasoning implies that a lifeless Solar System would place the Great Filter the farthest in our past, as Bostrom (2008) provocatively claimed: "I hope that our Mars probes discover nothing. It would be good news if we find Mars to be sterile. Dead rocks and lifeless sands would lift my spirit." Evidence that life never took hold elsewhere in the Solar System would increase confidence that the Great Filter is at or near the origin of life, although this might also imply that Earth is unique or rare as an inhabited planet. Bostrom (2008) noted that the possibility of the galaxy teeming with technological life that we are unable to detect is unlikely, as "if extraterrestrials do exist in any numbers, at least one species would have already expanded throughout the galaxy, or beyond."

While the original Great Filter argument presented by Hanson (1998) is suggestive of a single inhibitory step on the track from abiogenesis to galactic settlement, in principle there may be multiple low probability steps. The number of so-called "hard-steps" identified by authors varies, but tends to converge between four and six (Carter 2008; Watson 2008; McCabe & Lucas 2010; Lingam & Loeb 2019a). Accordingly, any failure to detect microbial life in the Solar System would not necessarily rule out an additional filter between abiogenesis and the rise of intelligence, or intelligence and galactic settlement, for example. However, whether there are multiple hard steps or only one, the Great Filter argument developed by Hanson (1998) and Bostrom (2008) suggests that the detection of life elsewhere is "worse news" the closer it is in developmental stage to human civilization, as such a discovery would enhance the probability that the hardest steps are in our future.

It also remains possible that technological civilizations, even those far more advanced than humans today, are unable to engage in galactic-scale settlement. This could be due to technological factors that inhibit interstellar travel, economic factors that render settlement cost-prohibitive, political factors that prevent effective governance over the requisite timescales, moral factors that inhibit societal longevity; biological factors that contribute to genetic degradation, or physical factors that place limits upon the rate of expansion (e.g., von Hoerner 1975; Finney & Jones 1986; Newman & Sagan 1981; Slobodian 2015; Mullan & Haqq-Misra 2019). Following the logic of the Great Filter argument, any such impediments to galactic-scale settlement must apply to technological life in all planetary systems; otherwise, a single exception would lead to a visibly settled galaxy. If interstellar travel and thus galactic-scale settlement is impossible, then even the most advanced extraterrestrial civilizations would be physically constrained to the environment of their planet or stellar system. In such a scenario, we would not expect widespread signs of extraterrestrial visitors, but we could still search for evidence of extraterrestrial civilizations in the form of spectroscopic technosignatures. Alternatively, some authors have suggested scenarios that allow wide scale galactic settlement that leave some planets (including Earth) apparently untouched (e.g., Landis 1998; Carroll-Nellenback et al. 2019; Đošović et al. 2019), thereby invalidating an underlying assumption of the Great Silence, although we may still expect remotely detectable evidence of their existence elsewhere via technosignatures.

Biosignatures and technosignatures both reveal large-scale processes that result from the interaction of life and technology with the planetary system. Just as the search for biosignatures is a systematic attempt to discover extraterrestrial biospheres, the search for technosignatures is concerned with identifying "technospheres" that result from observational signals of planetary-scale technology. As a way of framing the search for life and technology from the perspective of planetary evolution, Frank et al. (2017) developed a five-tier classification scheme based upon the forms of free energy generated on a planet. Under this classification scheme, Class I planets are close to radiative equilibrium and lack an atmosphere, such as Mercury. Class II planets retain an atmosphere and may therefore exhibit kinetic energy dissipation associated with large-scale circulation as well as some chemical free energy dissipation, with Mars as an example. Class III planets are inhabited with a thin biosphere, perhaps analogous to Archean Earth, where life does not exert strong feedbacks upon the planet. Class IV planets are inhabited and include stronger feedbacks between the biosphere and planetary system, with Earth after the Great Oxidation Event serving as an example. Extending one step further, Class V planets feature a thick biosphere plus dissipation from energy-intensive global-scale technology. Earth today resides as a "hybrid planet" between Class IV and Class V (Frank et al. 2017), with an emerging technosphere that is dimly detectable and an exponentially growing demand for energy. Humanity's successful transition to a Class V civilization will depend upon achieving a sustainable equilibrium between the availability and consumption of renewable energy sources.

Returning to the Great Filter, the abundance of planets in each of the five classes can be resolved through the spectroscopic characterization of exoplanets. Class I and II planets are already widely known inside and outside the solar system, but the abundance of Class III and IV planets could be constrained by searching for biosignatures. Similarly, a comprehensive search for spectroscopic technosignatures on any planets with observed biosignatures would place constraints on the abundance of Class V planets. The discovery that inhabited Class III and IV planets are commonplace but technological Class V planets are rare would likely place the Great Filter in our future, with no indication that our energy-intensive trajectory can be managed. Conversely, the discovery of widespread Class V planets would be genuine call for celebration, as this would be evidence that numerous inhabited exoplanets have successfully navigated the transition from energy-intensive activities analogous to our hybrid planet into a sustainable planetary-scale technosphere. Searching for evidence of biosignatures and technosignatures remains an important goal for understanding the future prospects of technological life on Earth.

## 3    Observing Habitable Planets

Suppose as a thought experiment that a mission like LUVOIR is launched in the near future and successfully characterizes a statistically significant sample of planets during its mission lifetime. In order to constrain the frequency of a planetary property, $\eta_x$, LUVOIR must observe a number of exoplanet candidates, $N_{ec}$, that reveal evidence of feature $x$. Assuming a binomial distribution for the probability of observing feature $x$ on a given exoplanet and a high-efficiency detector, the number of required exoplanet candidates is approximately equal to:

$$N_{ec} = \frac{\log(1-c)}{\log(1-\eta_x)} \tag{1}$$

where $c$ is the confidence of the constraint (Stark et al. 2014). In this article, both Class III and IV planets are denoted by $\eta_{life}$, the fraction of planets that develop life. For example, if $\eta_{life} = 0.05$, then LUVOIR can constrain the prevalence of inhabited planets (through the presence of spectral biosignatures) with 95% confidence by observing ~60 exoplanet candidates. Or to put it another way, if LUVOIR observes ~60 exoplanet candidates and finds no biosignatures on any of them, then this places a 95% confidence upper limit of $\eta_{life} \leq 0.05$ (Figure 1, left panel).

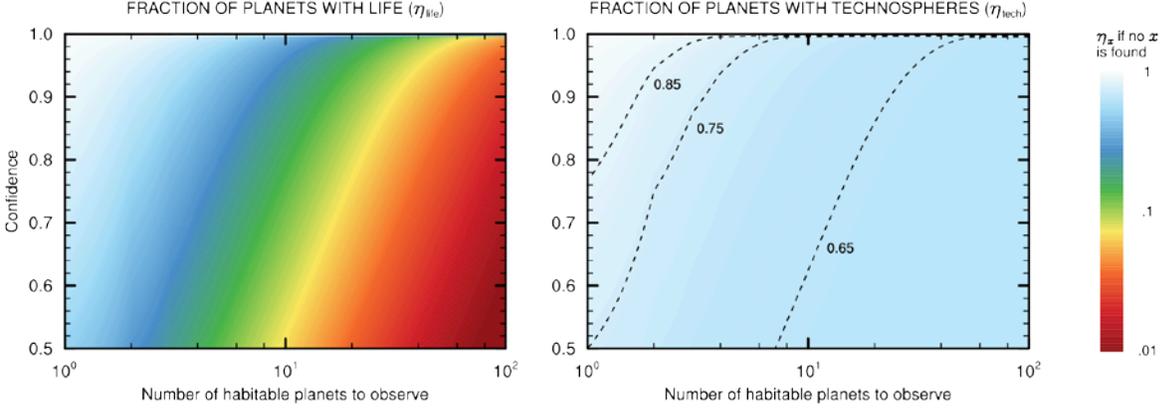

**Figure 1:** A mission such as LUVOIR that observes $N_{ec}$ habitable planets and finds no biosignatures on any of them will place an upper limit on the fraction of planets with life ($\eta_{life}$), given a confidence (left). Similarly, a comprehensive search for technosignatures with TSIG that finds none will place an upper limit on the fraction of inhabited planets with observable technospheres ($\eta_{tech}$), given a confidence (right). For example, a search with $N_{ec} = 60$ at 95% confidence that finds no biosignatures or technosignatures would place limits of $\eta_{life} \leq 0.05$ and $\eta_{tech} \leq 0.64$.

Suppose further that a second mission is designed with the intention of searching known exoplanet candidates for technosignatures. LUVOIR is not optimized for technosignatures at mid-infrared wavelengths and thus can only perform cursory searches for evidence of technological life at UV, visible, and near-infrared wavelengths (e.g., Lingam & Loeb 2017, Socas-Navarro 2018). The architecture of a mission intended to search for technosignatures might extend observations into the far infrared to detect thermal waste emissions and greenhouse gas absorption by complex molecules, but further details are beyond the scope of this article. For the present thought experiment, the hypothetical concept of a mission optimized to search for spectral technosignatures will be referred to as TSIG (TechnoSIGnature Surveyor). The LUVOIR mission is intended to be serviceable, so some of the capabilities of TSIG theoretically could be implemented into an extended LUVOIR mission. Another mission concept that could search for technosignatures is the ESA Large Interferometer for Exoplanets (LIFE), which would directly image nearby planetary systems at mid-infrared wavelengths (Defrère et al. 2018). LIFE may be the mission concept today most comparable to the hypothetical TSIG, although identifying the optimal range of wavelengths to search for technosignatures remains an open area of study. Ultimately, the cost-benefit of the addition of technosignature search capabilities will need to be determined for either a hybrid or dedicated mission (e.g., Lingam & Loeb, 2019b), but we cannot finalize these estimates before technosignature science reaches a higher level of maturity.

The quantity $\eta_{tech}$ is defined as the fraction of inhabited planets that show observable technospheres (Class V planets), so the number of habitable planets that TSIG must observe in order to determine the prevalence of technology in the galaxy is:

$$N_{tech} = \frac{\log(1-c)}{\eta_{life}\log(1-\eta_{tech})} \qquad (2)$$

If a search by TSIG observes the same ~60 exoplanet candidates as LUVOIR (which found $\eta_{life} \leq$ 0.05) and finds no technosignatures, then this would place a 95% confidence upper limit of $\eta_{tech} \leq$ 0.64 (Figure 1, right panel). While this is a relatively loose constraint on $\eta_{tech}$, it would demonstrate that many life-bearing planets have not transitioned into technospheres. Setting tighter constraints on $\eta_{tech}$ will require a more ambitious survey. At the other extreme, if LUVOIR finds that life is everywhere ($\eta_{life} = 1$) in a 60-planet sample, then a search of these same targets by TSIG that finds no technosignatures would place a 95% confidence upper limit of $\eta_{tech} \leq 0.05$. A full range of values for $N_{tech}$ are plotted in Figure 2 with $\eta_{life}$ and $\eta_{tech}$ between 0.01 and 1.0 and $c = 0.95$.

Class I and II planets are ubiquitous today, with examples inside and outside the solar system, but the discovery of even a single Class III or IV exoplanet would represent the monumental discovery that our biosphere is not alone. The discovery of abundant Class IV planets in particular would provide evidence that productive biospheres amenable for development of complex and intelligent life are also common. (Catling et al. 2005 suggested that an oxygenated atmosphere may even be a requirement for complex life, which would imply that Class IV planets should feature atmospheric $O_2$. While some researchers have proposed abiotic processes that could produce detectable molecular oxygen in planetary atmospheres, extensive strategies for fingerprinting or excluding abiotic origins of $O_2$ have likewise been developed; see Meadows et al. (2018) for a review). If a survey of a large sample of habitable planets by LUVOIR reveals that $\eta_{life}$ is large, then this would indicate that the origin of life is not necessarily a difficult step in planetary evolution. Even more optimistically, if follow-up searches for technosignatures by either LUVOIR or the more capable TSIG reveal that $\eta_{tech}$ is also large, then this would give confidence that the Great Filter does not exist, or it is in our past. If $\eta_{life}$ and $\eta_{tech}$ are large, then it would require ~50 detections of habitable terrestrial exoplanets with LUVOIR and TSIG to find evidence of a Class V planet (Figure 2).

Continuing the thought experiment, suppose now that LUVOIR finds $\eta_{life}$ is large but subsequent intensive searches by TSIG reveal $\eta_{tech} \sim 0$. The discovery that life is common but technology is rare would increase confidence that the Great Filter is near—either in our most recent evolutionary past or, most worryingly, in our immediate future. If Class III planets are found to be much more common than Class IV planets, then this would indicate that the Great Filter resides between the origin of life and the development of a complex biosphere. Such a scenario would indicate that Earth is one of the few (if only) planets to survive the Great Filter; however, we could not rule out a second "Greater Filter" awaiting us in the future. Alternatively, if Class III planets are vastly outnumbered by Class IV planets, but Class V planets are not found, then Earth may face substantial challenges in evolving its technosphere. This conclusion may seem bleak for the future of humanity, but perhaps it would also be inspiring. The discovery that most planets fail to transition to Class V could motivate our civilization to be the first. Conversely, such optimism could be displaced by the assumption of mediocrity and thus the lack of collective action toward solving major global problems. The detection of large $\eta_{life}$ and small $\eta_{tech}$ would also be consistent with the phase transition hypothesis (Annis 1999; Ćirković & Vukotić 2008), which would indicate that physical properties of the galaxy or universe have only recently enabled the development of Class V planets. Detecting large $\eta_{life}$ and small $\eta_{tech}$ would require ~100 observations of habitable terrestrial exoplanets with LUVOIR (Figure 2).

If LUVOIR detects very few biosignatures and finds that $\eta_{life}$ is small, then $\eta_{tech} \sim 0$ would suggest that Earth is alone as a hybrid planet. A lack of Class III, IV, and V planets could indicate that Earth is among the first in the galaxy to embark on this course of planetary evolution. This likewise could be encouraging or discouraging, as Earth's trajectory toward a Class V planet represents uncharted territory. But this possibility still leaves the future open to human settlement of the galaxy, although a Great Filter could still be waiting to challenge us toward this end. Confidently detecting small $\eta_{life}$ and $\eta_{tech}$ would hypothetically require ~3000 observations of habitable terrestrial exoplanets with both LUVOIR, its successor(s), and TSIG (Figure 2).

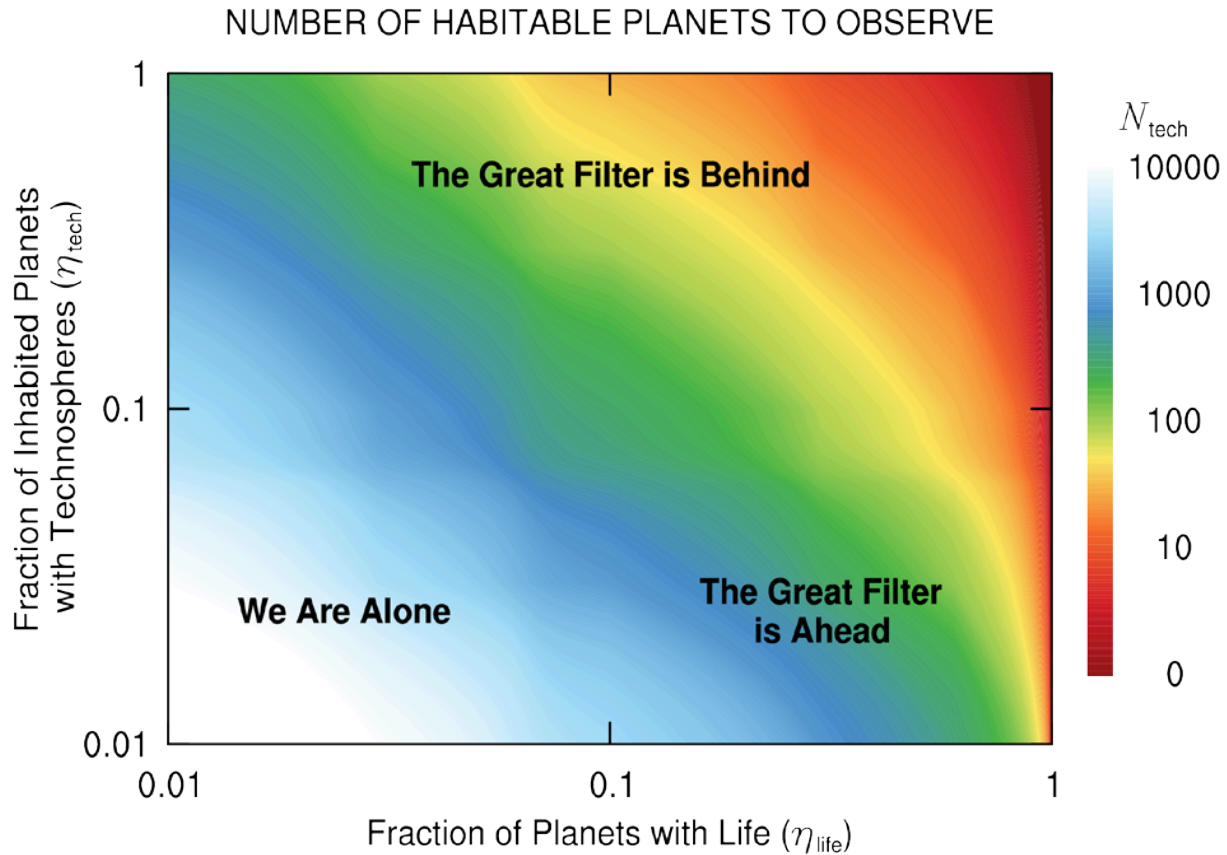

**Figure 2:** The number of habitable planet observations required to constrain the Great Filter (with 95% confidence) depends upon the fraction of planets with life ($\eta_{life}$) and the fraction of inhabited planets with observable technospheres ($\eta_{tech}$). A mission such as LUVOIR may able to constrain the value of $\eta_{life}$, but subsequent searches for technosignatures will be required to determine whether the Great Filter is in our past ($\eta_{tech} \geq \eta_{life}$) or future ($\eta_{tech} < \eta_{life}$).

It is also conceivable that $\eta_{life}$ is small, but $\eta_{tech}$ is still large. If life is rare but most planets with biospheres are Class V, then this would give confidence that the Great Filter is in our past at the origin of life. Observing small $\eta_{life}$ but large $\eta_{tech}$ would suggest two possibilities for the future of human civilization. The first interpretation would be that nearly all planets that independently evolve life transition into Class V planets. This would suggest that Earth's status as a hybrid planet between Class IV and Class V is a common transition, which would provide us with the assurance that the daunting challenges facing human civilization have been solved by other energy-intensive civilizations before us. This would also suggest that interstellar travel is impossible, so that human civilization is one of many technological civilizations in the galaxy bound to their home planetary system. But a second interpretation could be that biogenesis is extremely rare but interstellar proliferation of biospheres is commonplace, so that life on the numerous Class V planets all originate from a single ancestor planet. This would suggest that interstellar travel is possible and humans co-inhabit the galaxy with a handful civilizations that have already settled many planetary systems; however, this would not provide any greater confidence that human civilization will necessarily achieve a similar interstellar spacefaring status. Detecting small $\eta_{life}$ and large $\eta_{tech}$ would hypothetically require ~300 unique observations of habitable terrestrial exoplanets with both LUVOIR, its successor(s), and TSIG (Figure 2).

A final possibility deserves mention for any scenarios with $\eta_{tech} \sim 0$. Even if biospheres are common but technology appears to be absent, this could indicate that the galaxy is teeming with life but we have no way of recognizing extraterrestrial technology. For example, Grinspoon (2003) suggested that the galaxy is filled with "quasi-immortals" that have achieved long-term energy independence and sustainability as a planetary civilization but would be unrecognizable to us. Tarter et al. (2018) similarly suggested that technosignatures may masquerade as biosignatures given a civilization that has achieved a sufficient level of sustainability. Yet another possibility is that suspected technosignatures may actually originate from non-technological biological processes (Raup 1992), which would confound the unambiguous identification of extraterrestrial technology. Smart (2012) further speculated that universal intelligence may "transcend" into a black-hole-like environment that would elude most attempts at characterization. Even with our most imaginative and comprehensive searches, it remains possible that Class V planets would be so drastically different (perhaps blending invisibly into their environments) that they could be extremely commonplace but impossible for us to detect. However, without any observational constraints on $\eta_{tech}$, we still could not rule out the possibility that a Great Filter awaits between Earth's present state and our possible future as an undetectable Class V planet.

## 4    The Search for Technosignatures

The search for technosignatures is an important science goal with significant philosophical and political implications as humanity proceeds into the future. Theoretical frameworks currently exist from decades of research into spectral characterization of biosignatures that would reveal the abundance of Class I, II, III, and IV planets. Any comprehensive search for spectral biosignatures should be corroborated by a similarly comprehensive search for technosignatures at any and all wavelengths. Clearly defining and identifying the tell-tale spectral features of Class V planets will require advancing the theoretical and observational science of technosignatures to the same level of sophistication as biosignature science today. Even more so than biosignature research, developing a library of theoretical expectations for plausible technosignatures will require transdisciplinary collaboration between a broad range of scholars.

If optimism prevails and we find that Class V planets are commonplace, placing the Great Filter in our past, then we can breathe easy and look forward to a future of studying (and perhaps communicating with) our interstellar neighbors. But if the search for biosignatures reveals that life is everywhere while technology is not, then our challenge is even greater to secure a sustainable future.

**Acknowledgements**
This article was inspired by the sessions on "The Astrobiology of the Anthropocene" and "Searching for Technosignatures" at the 2019 Astrobiology Science Conference in Bellevue, Washington. This work benefited from the authors' participation in the NASA Nexus for Exoplanet System Science (NExSS) research coordination network under award NNH13DA017C and insight from the 2018 NExSS Technosignature Workshop Report. J.H. and R.K. gratefully acknowledge funding from the NASA Habitable Worlds program under award 80NSSC17K0741 and the Virtual Planetary Laboratory under NASA award 80NSSC18K0829. R.K. also acknowledges funding from the Sellers Exoplanet Environments Collaboration. Any opinions, findings, and conclusions or recommendations expressed in this material are those of the authors and do not necessarily reflect the views of NASA.